\documentclass[aps,tightenlines,
english,twocolumn,preprintnumbers,amsmath,amssymb,nofootinbib,superscriptaddress]{revtex4-1}
\usepackage{graphics,bm}
\usepackage{epsfig}
\usepackage{graphicx}
\usepackage{amsmath}
\usepackage{wrapfig}
\usepackage{color}

\newcommand{\os}{\text{\tiny OS}}
\newcommand{\ms}{\overline{\text{\tiny MS}}}

\newcommand{\beq}{\begin{equation}}
\newcommand{\eeq}{\end{equation}}
\newcommand{\bqa}{\begin{eqnarray}}
\newcommand{\eqa}{\end{eqnarray}}

\def\sumint{\hbox{$\sum$}\!\!\!\!\!\!\int}
\def\square{\vcenter{\vbox{\hrule height.4pt
          \hbox{\vrule width.4pt height4pt
          \kern4pt\vrule width.3pt}\hrule height.4pt}}}

\begin{document}


\title{Inhomogeneous phases at finite density in an external
magnetic field}

\author{Jens O. Andersen}
\email{andersen@tf.phys.ntnu.no}
\affiliation{Department of Physics, Faculty of Natural Sciences,NTNU, 
Norwegian University of Science and Technology, H{\o}gskoleringen 5,
N-7491 Trondheim, Norway}
\author{Patrick Kneschke}
\email{patrick.kneschke@uis.no}
\affiliation{Faculty of Science and Technology, University of Stavanger,
N-4036 Stavanger, Norway}
\date{\today}

\begin{abstract}
The two-flavor quark-meson model is used as a low-energy effective model for
QCD to study inhomogeneous chiral
condensates at finite quark chemical potential $\mu$
in a constant magnetic background $B$. We determine the
parameters of the model by matching the meson and quark
masses, and the pion decay constant to their physical values using the on-shell and 
modified minimal subtraction schemes. 
We calculate the free energy in the mean-field approximation for a chiral
density wave using dimensional regularization.
The system
has a surprisingly rich phase structure.

\end{abstract}

\maketitle

\section{Introduction}
QCD at finite temperature and baryon density has been studied in detail
for several decades, largely spurred by its relevance for the early universe,
heavy-ion collisions, and compact stars \cite{revalford,revhatsuda}.
At zero baryon density, one can perform lattice simulations
to calculate the thermodynamic functions and the transition temperature
associated with chiral symmetry restoration and deconfinement.
For physical quark masses and two flavors, the transition is a crossover at a
temperature of around 155 MeV \cite{latt1,latt2,latt3,latt4}. 
At finite $\mu_B$,  the infamous sign
problem prevents the use of standard Monte Carlo techniques based
on importance sampling. This implies that for large parts of the
phase diagram, we have to rely on model calculations.
A drawback of model calculations is that some of the predictions
are not robust; for example, the existence of certain
phases may depend on the values of its parameters.
Only 
at asymptotically high densities are we confident about
the phase and the properties of QCD. In this limit, the
ground state of QCD is the color-flavor-locked  phase
which is a color superconducting phase.

The Nambu-Jona-Lasinio (NJL) model and the quark-meson (QM) model
are examples of low-energy models of QCD; they share some of 
its properties such as chiral symmetry and the breaking of it in the vacuum.
While these models incorporate the chiral aspects of QCD very well, they
are not confining. This has led to the introduction of the Polyakov
loop, which is an (approximate) order parameter for 
confinement \cite{svet}.
By coupling an $SU(N_c)$ background gauge field $A_{\mu}$
to the chiral model one can mimic confinement in
QCD in a statistical sense \cite{fukk}.

There are other control parameters in addition to the temperature $T$
and the baryon chemical potential $\mu_B$, for example an external
magnetic field $B$. See e.g. 
\cite{ebert2000,inag,manuel,jorge,harmen,prov1,frolov,allen,grun,sponmag,cao,nishi,lif} for model 
calculations.
QCD in a magnetic background at $\mu_B=0$ is free of the sign problem 
and consequently one can perform lattice simulations.
Lattice simulations of QCD in a constant magnetic background
have been carried out in recent years to study the 
chiral condensate as a function of $T$ and to calculate
the transition temperatures for chiral symmetry restoration and
deconfinement \cite{delia,bali1,bali2,endrodi1}.          
However, for finite $\mu_B$, the sign problem again prevents one from
using Monte Carlo techniques. 
\\ \indent
QCD at finite $\mu_B$ and $B$
provides an example of where naive model calculations may go wrong.
Using chiral perturbation theory and including the 
Wess-Zumino-Witten term, it has been shown that the ground state
of QCD for certain values of $\mu_B$ and $B$
is a spatially modulated condensate of neutral pions \cite{son}.
Recently, it has been shown that the exact solution 
is a chiral soliton lattice \cite{brauner}. The importance of this
results is that it is a model-independent statement and therefore
robust. The critical magnetic field $B_{\rm crit}(\mu_B)$
has also been estimated and for
$\mu_B\approx900$ MeV, which corresponds to the onset of nuclear matter,
it is approximately $10^{19}$ Gauss, which may be found in the core
of magnetars. This corresponds to $|eB|=1.88 m_{\pi}^2$.
\\ \indent
Calculating the thermodynamic potential in these models, one encounters 
ultraviolet divergences due to vacuum fluctuations. In NJL-model calculations, 
one often uses a sharp
three-dimensional cutoff to regulate them \cite{klev}. 
However, in the case of inhomogeneous
condensates, this typically leads to artifacts of the regularized free energy. 
For example,
the chiral density wave is characterized by the modulus $\Delta$ and 
the wave vector $q$. In the
limit $\Delta\rightarrow0$, the 
free energy cannot depend on the wave vector. If this is the case,
one must find suitable terms to subtract from  the free energy
to make it well defined \cite{tomas}. 
The problem with a three-dimensional cutoff
is that there is an asymmetry in the values of the energies that
contributes to the free energy from the different branches of
the quasiparticle branches. One can avoid this artifact in most
cases \footnote{A counterexample is the NJL model in 1+1 dimensions
with an isospin chemical potential \cite{symren}, where one must
add a term $-{N_c\over\pi}\mu_I^2$.} 
by introducing a {\it symmetric energy cutoff}
\cite{frolov,symren}, i.e. one cuts off the 
energy rather than the three-momentum.
A convenient alternative to imposing sharp cutoffs is dimensional 
regularization. In Ref. \cite{jens1}, it was shown how one can use
dimensional regularization to calculate the vacuum energy
in various models for a chiral-density wave. In some cases, one still
has to subtract suitable terms to make the vacuum energy well defined
in the limit $\Delta\rightarrow0$. In the present case, no 
such terms are necessary to subtract, as we will show.

The article is organized as follows. In the next section, we briefly
discuss the quark-meson model. The free energy is calculated
in the mean-field approximation. In section \ref{results}
we present our results.
First we discuss the case of a homogeneous chiral condensate and map out the
phase diagram. We next consider the
chiral density wave and inhomogeneous phases. 
Details of the calculations can be found in two appendices.

\section{Quark-meson model and free energy}
\label{qm}
\subsection{Quark-meson model}
The Euclidean Lagrangian of the two-flavor quark-meson model is
\bqa\nonumber
{\cal L}&=&
{1\over2}\left[(\partial_{\mu}\sigma)^2
+(\partial_{\mu}{\boldsymbol \pi})^2\right]
+{1\over2}m^2(\sigma^2+{\boldsymbol\pi}^2)
\\ \nonumber
\nonumber
&&+{\lambda\over24}(\sigma^2+{\boldsymbol\pi}^2)^2
-h\sigma
\\&&
+\bar{\psi}\left[/\!\!\!\!D-
(\mu+{1\over2}\tau_3\mu_I)
\gamma^0
+g(\sigma+i{\gamma^5\boldsymbol\tau}\cdot{\boldsymbol\pi})\right]\psi
\;,
\label{lag}
\eqa
where $\psi$ is a flavor doublet as well as an $SU(N_c)$-plet,
\bqa
\psi&=&
\left(
\begin{array}{c}
u\\
d
\end{array}\right)
\eqa
Moreover,
$\mu_B=3\mu={3\over2}(\mu_u+\mu_d)$ and $\mu_I={1\over2}(\mu_u-\mu_d)$
are the baryon and isospin 
chemical potentials in terms of the quark chemical potentials
$\mu_u$ and $\mu_d$, 
and $\tau_i$ ($i=1,2,3$) are the Pauli matrices in flavor space.
$D_{\mu}=\partial_{\mu}-q_fA_{\mu}$ is the covariant derivative
and $q_f$ is the electric charge of flavor $f=u,d$. 
In the case of a constant magnetic field
$B$ directed along the $z$-axis, we choose the gauge 
$A_{\mu}=(0,0,-Bx,0)$. 
In addition to the global $SU(N_c)$ symmetry, the 
Lagrangian~(\ref{lag}) 
has, for $A_{\mu}=0$,
a $U(1)_B\times SU(2)_L\times SU(2)_R$ symmetry for 
$h=0$ and $U(1)_B\times SU(2)_V$ for $h\neq0$. 
When $\mu_u\neq\mu_d$, this symmetry is reduced to 
$U(1)_B\times U(1)_{I_3L}\times U(1)_{I_3R}$ for $h=0$ and
$U(1)_B\times U(1)_{I_3}$ for $h\neq0$. 
The constant magnetic field breaks Lorentz invariance;
it also breaks the to $SU(2)_V$ down to $U(1)_V$.
In the remainder of the paper, we choose $\mu_u=\mu_d$
and $h\neq0$.
In  the  vacuum,  the $\sigma$ field  acquires  a  nonzero  
vacuum expectation value, which is denoted by
$\phi_0$.  In order to study inhomogeneous phases,  we make an
ansatz  for  the  space-time  dependence  of  the  mesonic
mean fields.   In  the  literature,  mainly  one-dimensional
modulations  have  been  considered,  for  example  chiral-density waves 
(CDW) and soliton lattices \cite{buballarev}.  We
opt  for  the  simplest, namely a one-dimensional chiral-density wave,
although this might not be the modulation with the lowest energy.
The ansatz is
\bqa
\phi_0(z)={\Delta\over g}\cos(qz)
\;,
\hspace{0.5cm}
\pi(z)={\Delta\over g}\sin(qz)\;,
\label{ansatz}
\eqa
where $\Delta$ and $q$ are real parameters.
The bilinear term in the Lagrangian in Minkowski space is
\bqa
\bar{\psi}\left[
iD\!\!\!\!/+\gamma_0\mu-\Delta e^{i\tau_3qz}
\right]\psi\;.
\eqa
We next redefine the quark fields, 
$\psi\rightarrow e^{-{1\over2}i\gamma_5\tau_3qz}\psi$ and
$\bar{\psi}\rightarrow\bar{\psi}e^{-{1\over2}i\gamma_5\tau_3qz}$.
This corresponds to a unitary transformation of the Dirac Hamiltonian
${\cal H}\rightarrow{\cal H}^{\prime}
=e^{{1\over2}i\gamma_5\tau_3qz}{\cal H}e^{-{1\over2}i\gamma_5\tau_3qz}$.
As pointed out in Ref. \cite{frolov}, a field redefinition as the one above
requires extra care in the presence of background gauge fields as 
a change of the path-integral measure must be taken into account.
Using the method of Fujikawa \cite{fuji}, there is an extra factor of
$e^{\epsilon^{\alpha\beta\mu\nu}F_{\alpha\beta}F_{\mu\nu}}$.
In a constant magnetic field, the
term ${\epsilon^{\alpha\beta\mu\nu}}F_{\alpha\beta}F_{\mu\nu}$ vanishes
and the measure is invariant.
The Dirac operator can be written as
\bqa
D&=&[iD\!\!\!\!/+\gamma_0\mu-\Delta +\mbox{$1\over2$}\gamma_5\gamma_3\tau_3q
]\;.
\eqa
The spectrum is given by~\cite{gorbar}
\bqa
E_{\pm}^2&=&
\left(\sqrt{p_{z}^2+\Delta^2}\pm\mbox{$q\over2$}\right)^2
+|q_fB|(2k+1-\sigma_z)\;,
\label{disprel}
\eqa
where $q_f$ is the charge of the quark flavor,
$\sigma_z=\pm1$, and $k=0,1,2...$. 
We note that  the lowest Landau level, which corresponds to 
$\sigma_z=1$ and $k=0$ is independent of the magnetic field.

At tree level, the relations between the parameters 
$m^2$, $\lambda$, $g^2$, and $h$ of the Lagrangian
Eq. (\ref{lag}) and the physical observables $m_\sigma$, $m_\pi$, $m_q$,
and $f_\pi$ are
\bqa
\label{tr1}
&&m^2=-{1\over2}\left(m_{\sigma}^2-3m_{\pi}^2\right)\;,\;
\lambda=3{(m_{\sigma}^2-m_{\pi}^2)\over f_{\pi}^2}\;,\;
\\ &&
g^2={m_q^2\over f_{\pi}^2}
\;,\;\;
h=m_\pi^2f_\pi
\;.
\label{tr4}
\eqa
We also have that $f_{\pi}$ minimizes the
tree-level potential in the vacuum,
$V={1\over2}m^2\phi_0^2+{\lambda\over24}\phi_0^4$.

Expressed in terms of
physical quantities, the tree-level potential is 
\bqa\nonumber
V_{\rm tree}&=&
{1\over2}B^2
+
{1\over2}\left({q\Delta\over g}\right)^2     
+{1\over2}m^2{\Delta^2\over g^2}
+{\lambda\over24}{\Delta^4\over g^4}
-h{\Delta\over g}
\\ \nonumber
&=&
{1\over2}B^2
+
{1\over2}f_\pi^2q^2{\Delta^2\over m_q^2}
-{1\over4}f_\pi^2(m_{\sigma}^2-3m_{\pi}^2){\Delta^2\over m_q^2}
\\ &&
+{1\over8}f_\pi^2(m_{\sigma}^2-m_{\pi}^2){\Delta^4\over m_q^4}
-m_\pi^2f_\pi^2{\Delta\over m_q}\;.
\label{tri}
\eqa
\subsection{Free energy}
The free energy is calculated in the mean-field approximation, where we
treat the bosonic degrees of freedom at tree level.

The one-loop contribution to the effective potential from the fermions is
\bqa
V_1&=&-N_c\sum_{f,\pm}\sumint_{\{P\}}^B\log\left[
P_0^2+E_{\pm}^2
\right]\;,
\eqa
where the sum-integral is defined by
\bqa
\sumint_{\{P\}}^B&=&
{|q_fB|\over2\pi}T\sum_{\sigma_z=\pm1}\sum_{k=0}^{\infty}\sum_{P_0}\int_{p_z}
\;.
\label{sumint}
\eqa
The sum is over Matsubara frequencies $P_0=(2n+1)\pi T$ ($n=,0,\pm1,\pm2...$),
and Landau levels $k$ and $\sigma_z$. The integral is
a dimensionally regulated integral in $d-2=1-2\epsilon$ dimensions, 
which is defined by
\bqa
\int_{p_z}&=&
\left({e^{\gamma_E}\Lambda\over4\pi}\right)^{\epsilon}\int{d^{d-2}p\over(2\pi)^{d-2}}
\;.
\label{definte}
\eqa
where $\Lambda$ is the renormalization scale in the
modified mini\-mal subtraction scheme $\overline{\rm MS}$.
Summing of the Matsubara frequencies $P_0$, we obtain
\bqa\nonumber
V_{1}&=&-N_c\sum_{f,\pm,k,\sigma_z}{|q_fB|\over4\pi}
\int_{p_z}
\bigg\{E_{\pm}+
T\log\left[1+e^{-\beta(E_{\pm}-\mu)}\right]
\\&&
+T\log\left[1+e^{-\beta(E_{\pm}+\mu)}
\right]\bigg\}
\;.
\label{vacsum}
\eqa
The vacuum part of the free energy is
\bqa
V_{\rm vac}^B&=&-N_c\sum_{f,\pm,k\sigma_z}{|q_fB|\over4\pi}
\int_{p_z}E_{\pm}
\label{vacb}
\;.
\eqa
Eq. (\ref{vacb}) is ultraviolet divergent. 
In dimensional regularization the power divergences
are set to zero and the logarithmic divergences show up as poles in 
$\epsilon$. All the poles, except one which is proportional to
$(q_fB)^2$, are identical to the poles found when evaluating the vacuum
energy $V_{\rm vac}$ for $B=0$.
We can therefore isolate
these divergences by adding and subtracting $V_{\rm vac}$
The difference $V_{\rm vac}^B-V_{\rm vac}$ then contains
a pole in epsilon which is proportional to $(q_fB)^2$. This 
divergence can be then be isolated
by adding and subtracting a second divergent term.
All the divergences are subsequently eliminated by minimal subtraction.

The integral representation of the energy difference $E_{\pm}-E_{0\pm}$
\bqa
E_{\pm}-E_{0\pm}&=&-{1\over(4\pi)^{1\over2}}\int_0^{\infty}
{ds\over s^{3\over2}}
\left[e^{-s E_{\pm}^2}-e^{-s E_{0\pm}^2}\right]\;,
\label{sving1}
\eqa
where $E_{0\pm}$ is the dispersion relation in the case $B=0$, 
given by
\bqa
E_{0\pm}^2&=&
\left(\sqrt{p_{z}^2+\Delta^2}\pm\mbox{$q\over2$}\right)^2+p_{\perp}^2\;,
\eqa
where $p_{\perp}^2=p_x^2+p_y^2$. For notational convenience we write
the dispersion relations as
\bqa
E_{\pm}^2=a_{\pm}^2+M_B^2\;,
\hspace{1cm}
E_{0\pm}^2=a_{\pm}^2+p_{\perp}^2\;,
\eqa
where $a_{\pm}^2=(\sqrt{p_{z}^2+\Delta^2}\pm{q\over2})^2$
and $M_B^2=|q_fB|(2k+1-\sigma_z)$. 
Using Eq. (\ref{sving1}), the
vacuum energy density difference can be written as
\bqa\nonumber
V_{\rm vac}^B-V_{\rm vac}&=&
-{N_c\over(4\pi)^{1\over2}}
\sum_{f,\pm}
\int_{p}\int_0^{\infty}{ds\over s^{3\over2}}e^{-s(a_{\pm}^2+p_{\perp}^2)}\;,
\\ 
&&
\hspace{-2cm}
+{N_c\over(4\pi)^{3\over2}}
\sum_{f,\pm,k,\sigma_z}|q_fB|\int_{p_z}\int_0^{\infty}{ds\over s^{3\over2}}
e^{-s(a_{\pm}^2+M_B^2)}\;. 
\eqa
Integrating over $p_{\perp}$ directly in two dimensions in the first
term and 
summing over $k$ and $\sigma_z$ in the second term, we find
\begin{widetext}
\bqa
V_{\rm vac}^{B}-V_{\rm vac}
&=&
{N_c\over(4\pi)^{3\over2}}
\sum_{f,\pm}\int_{p_z}\int_0^{\infty}{ds\over s^{5\over2}}
\bigg[|q_fB|s
\coth(|q_fB|s)-1
\bigg]e^{-s a_{\pm}^2}\;.
\label{divint}
\eqa
\begin{figure}
\end{figure}
The integral in Eq. (\ref{divint}) is divergent for small $s$, i.e.
in the ultraviolet. Expanding the integrand in (\ref{divint}),
it is straightforward to see that the UV-divergence is canceled by
the term 
\bqa\nonumber
V_{\rm div}^B &=&{2N_c\over3(4\pi)^{3\over2}}
\sum_{f}\int_{p_z}\int_0^{\infty}{ds\over s^{5\over2}}e^{-2s|q_fB|}
(q_fBs)^2e^{-sp_z^2}
=
N_c\sum_f\left({e^{\gamma_E}\Lambda^2\over2|q_fB|}\right)^{\epsilon}
{2(q_fB)^2\over3(4\pi)^2}
\Gamma(\epsilon)
\\ 
&=&
N_c\sum_f\left({\Lambda^2\over2|q_fB|}\right)^{\epsilon}
{2(q_fB)^2\over3(4\pi)^2}\left[{1\over\epsilon}
+{\cal O}(\epsilon)
\right]\;,
\label{dzz}
\eqa
where the extra exponential factor
$e^{-2s|q_fB|}$
ensures that the integral
is convergent in the 
infrared, i.e. for large values of $s$.
Subtracting Eq. (\ref{dzz}) from Eq. (\ref{divint}), we 
obtain the convergent result for  
the difference between the two vacuum energy densities 
\bqa
V_{\rm vac}^B-V_{\rm vac}&=&
{N_c\over(4\pi)^{3\over2}}\sum_{f,\pm}\int_{p_z}\int_0^{\infty}{ds\over s^{5\over2}}
\bigg\{\Big[
|q_fB|s\coth(|q_fB|s)-1\Big]e^{-sa_{\pm}^2}
-{1\over3}e^{-2s|q_fB|}(q_fBs)^2e^{-sp_z^2}
\bigg\}
\;.
\label{dir}
\eqa
\end{widetext}
The integral in Eq. (\ref{dir}) is 
is convergent in the ultraviolet and in the infrared. 
The integral over $p_z$
can therefore be evaluated in $d=1$ dimensions. 
For $\Delta=0$, this can be done explicitly.
Integrating over $p_z$ and summing
over $\pm$ yields
\bqa\nonumber
V_{\rm vac}^B-V_{\rm vac}&=&{2N_c\over(4\pi)^2}\sum_{f}
\int_0^{\infty}{ds\over s^3}
\Big[|q_fB|s\coth(|q_fB|s)
\\ &&-1-{1\over3}e^{-2s|q_fB|}(q_fBs)^2\Big]\;.
\eqa
This integral is independent of $q$ showing that
the vacuum energy is well defined.

We next consider the divergent part of the vacuum energy, which is
given by
\bqa
V_{\rm vac}&=&-2N_c\int_p\left(E_{0+}+E_{0-}\right)\;,
\label{v0}
\eqa
where the integral is analogous to Eq. (\ref{definte}), but now
in $d=3-2\epsilon$ dimensions.
Introducing the variable $u=\sqrt{p_z^2+\Delta^2}$
and integrating over angles in the $(p_x,p_x)$ plane, we can write
Eq. (\ref{v0}) as
\begin{widetext}
\bqa
V_{\rm vac}&=&
-{16 N_c(e^{\gamma_E}\Lambda^2)^{\epsilon}\over(4\pi)^{2}\Gamma(1-\epsilon)}
\sum_{\pm}
\int_{\Delta}^{\infty}
{u\,du\over\sqrt{u^2-\Delta^2}}
\int_{0}^{\infty}dp_{\perp}
\sqrt{\left(u\pm{q\over2}\right)^2+p_{\perp}^2}\,p_{\perp}^{1-2\epsilon}\;.
\label{integralane}
\eqa
The strategy is to isolate the ultraviolet divergences 
in Eq. (\ref{integralane}) by expanding the integrand in powers
of $q$
and identifying appropriate subtraction terms ${\rm sub}(u,p_{\perp})$.
Integrating the subtraction terms 
can be done in dimensional regularization, 
while the integral of $E_{\pm}-{\rm sub}(u,p_{\perp})$
is finite and can be calculated directly in three dimensions.
The subtraction terms ${\rm sub}(u,p_{\perp})$
is found by expanding Eq. (\ref{integralane}) through order $q^4$.
This yields
\bqa
{\rm sub}(u,p_{\perp})
&=&
2\sqrt{u^2+p^2_{\perp}}
+{q^2p_{\perp}^2\over4(u^2+p_{\perp}^2)^{3\over2}}
+{q^4p_{\perp}^2(4u^2-p_{\perp}^2)\over64(u^2+p_{\perp}^2)^{7\over2}}\;.
\label{sub22}
\eqa
We can then write $V_{\rm vac}=V_{\rm div}+V_{\rm fin}$, where
\bqa
V_{\rm div}&=&
-{16 N_c(e^{\gamma_E}\Lambda^2)^{\epsilon}\over(4\pi)^{2}\Gamma(1-\epsilon)}
\int_{\Delta}^{\infty}
{u\,du\over\sqrt{u^2-\Delta^2}}
\int_{0}^{\infty}dp_{\perp}
{\rm sub}(u,p_{\perp})p_{\perp}^{1-2\epsilon}
\;,
\\ 
V_{\rm fin}&=&
-{16 N_c(e^{\gamma_E}\Lambda^2)^{\epsilon}\over (4\pi)^{2}\Gamma(1-\epsilon)}
\sum_{\pm}
\int_{\Delta}^{\infty}
{u\,du\over\sqrt{u^2-\Delta^2}}
\int_{0}^{\infty}dp_{\perp}
\left[\sqrt{\left(u\pm{q\over2}\right)^2+p_{\perp}^2}
-{1\over2}{\rm sub}(u,p_{\perp})\right]
p_{\perp}^{1-2\epsilon}
\;.
\eqa
The integral $V_{\rm fin}$ can now be calculated directly in three dimensions.
After integrating over $p_{\perp}$, we find
\bqa
V_{\rm fin}&=&
-{16 N_c
\over3(4\pi)^{2}}
\int_{\Delta}^{\infty}
{u\,du\over\sqrt{u^2-\Delta^2}}(u-\mbox{$q\over2$})^2\left[
\left(u-\mbox{$q\over2$}\right)-\big|u-\mbox{$q\over{2}$}\big|
\right]\;.
\eqa
This finally yields
\bqa \nonumber
V_{\rm fin}
&=&
-{32 N_c\over3(4\pi)^2}\int_{\Delta}^{\infty}
{u\,du\over\sqrt{u^2-\Delta^2}}(u-\mbox{$q\over2$})^3
\theta(\mbox{$q\over2$}-\Delta)
\\ \nonumber
&=&
{N_c\over3(4\pi)^2}
\left[q\sqrt{{q^2\over4}-\Delta^2}(26\Delta^2+q^2)
-12\Delta^2(\Delta^2+q^2)\log{{q\over2}+\sqrt{{q^2\over4}-\Delta^2}\over\Delta}
\right]
\theta(\mbox{$q\over2$}-\Delta)
\\
&=&f(\Delta,q)
\;.
\label{finf}
\eqa
We next  integrate $V_{\rm div}$ using dimensional regularization.
Again this is done by first integrating over $p_{\perp}$ and then over $u$.
This yields
\bqa
V_{\rm div}&=&
{2N_c\over(4\pi)^2}
\left({e^{\gamma_E}\Lambda^2\over\Delta^2}\right)^{\epsilon}
\left[
2\Delta^4
\Gamma(-2+\epsilon)+q^2\Delta^2\Gamma(\epsilon)
+{q^4\over12}(-1+\epsilon)\Gamma(1+\epsilon)
\right]\;.
\label{vd}
\eqa
\end{widetext}
The divergent parts of the vacuum energy are given by 
Eqs. (\ref{dzz}) and (\ref{vd}), and require renormalization.
In the $\overline{\rm MS}$ scheme, the poles in $\epsilon$
are removed by
multiplying the $B^2$ term, the 
mass parameter and the couplings in the tree-level potential
(\ref{tri})
by $Z_A$,
$Z_{m^2}$, $Z_{\lambda}$, $Z_{g^2}$, and $Z_h$, respectively, where
\begin{widetext}
\begin{align}
Z_A&=1-N_c\sum_f{4q_f^2\over3(4\pi)^2\epsilon}\;,&
Z_{m^2}&=1+{4g^2N_c\over(4\pi)^2\epsilon}\;,&
Z_{\lambda}&=1+{8N_c\over(4\pi)^2\epsilon}\left[
\lambda g^2-6g^4\right]\;,&
\\ 
Z_{g^2}&=1+{4g^2N_c\over(4\pi)^2\epsilon}\;,
&
Z_h&=1+{2g^2N_c\over(4\pi)^2\epsilon}
\;.
\end{align}
After renormalization, the 
vacuum energy in the mean-field approximation reads
\bqa\nonumber
V&=&{1\over2}B_{\ms}^2+
{1\over2}q^2{\Delta^2\over g_{\ms}^2}
+{1\over2}m_{\ms}^2{\Delta^2\over g_{\ms}^2}
+{\lambda_{\ms}\over24}{\Delta^4\over g_{\ms}^4}
-h_{\ms}{\Delta\over g_{\ms}}
+{N_c}\sum_f{2(q_fB)^2\over3(4\pi)^2}\log{\Lambda^2\over2|q_fB|}
\\ && \nonumber
+{2N_c\Delta^2q^2\over(4\pi)^2}
\log{\Lambda^2\over\Delta^2}
+{2N_c\Delta^4\over(4\pi)^2}\left[
\log{\Lambda^2\over\Delta^2}+{3\over2}\right]
-{N_cq^4\over6(4\pi)^2}
+f(\Delta,\mbox{$q\over2$})
\\ && 
+
{N_c\over(4\pi)^{3\over2}}\sum_{f,\pm}\int_{p_z}\int_0^{\infty}{ds\over s^{5\over2}}
\bigg\{\big[
|q_fB|s\coth(|q_fB|s)-1\big]e^{-sa_{\pm}^2}
-{1\over3}e^{-2s|q_fB|}(q_fBs)^2e^{-sp_z^2}\bigg\}\;,
\label{potms}
\eqa
\end{widetext}
where the subscript $\ms$ indicates that the coupling are running.
The running field, mass and couplings constants satisfy 
the following renormalization group equations 
\bqa
\label{run0}
\Lambda{dB_{\ms}^2(\Lambda)\over d\Lambda}&=&
-N_c\sum_f{4q_f^2B_{\ms}^2(\Lambda)\over3(4\pi)^2}\;,
\\
\label{run1}
\Lambda{dm_{\ms}^2(\Lambda)\over d\Lambda}&=&
{8N_cm^2_{\ms}(\Lambda)g_{\ms}^2(\Lambda)
\over(4\pi)^2}\;,
\\
\Lambda{dg_{\ms}^2(\Lambda)\over d\Lambda}&=&{8N_cg_{\ms}^4(\Lambda)\over(4\pi)^2}
\;,
\\
\Lambda{d\lambda_{\ms}(\Lambda)\over d\Lambda}&=&{16N_c\over(4\pi)^2}
\left[\lambda_{\ms}(\Lambda) g^2_{\ms}(\Lambda)-6g^4_{\ms}(\Lambda)
\right]\;,
\\
\Lambda{dh_{\ms}(\Lambda)\over d\Lambda}&=&{4N_cg^2_{\ms}(\Lambda)
h_{\ms}(\Lambda)\over(4\pi)^2}
\label{run4}
\;.
\eqa
The solutions to Eqs. (\ref{run0})--(\ref{run4}) are
\bqa
\label{sol0}
B_{\ms}^2(\Lambda)&=&
{B_0^2\over1+\sum_f{4q_f^2B_0^2N_c\over3(4\pi)^2}
\log{\Lambda^2\over m_q^2}
}
\;, \\
\label{sol1}
m_{\ms}^2(\Lambda)&=&
{m_0^2\over1-{4g_0^2N_c\over(4\pi)^2}
\log{\Lambda^2\over \Lambda_0^2}
}\;.
\\
g_{\ms}^2(\Lambda)&=&
{g_0^2\over1-{4g_0^2N_c\over(4\pi)^2}
\log{\Lambda^2\over \Lambda_0^2}
}\;,
\\
\lambda_{\ms}(\Lambda)&=&{\lambda_0-{48g_0^4N_c\over(4\pi)^2}
\log{\Lambda^2\over \Lambda_0^2}
\over\left(1-{4g_0^2N_c\over(4\pi)^2}
\log{\Lambda^2\over \Lambda_0^2}
\right)^2}\;,
\label{sol3}
\\
h_{\ms}(\Lambda)&=&
{h_0\over1-{2g_0^2N_c\over(4\pi)^2}
\log{\Lambda^2\over \Lambda_0^2}
}\;,
\label{sol5}
\eqa
where $B_0$, $m_0^2$, $\lambda_0$, $g_0^2$, and $h_0$ are constants.
The parameters $m_0^2$, $\lambda_0$, $g_0^2$, and $h_0$, are the 
values of the running parameters at the scale $\Lambda_0$, 
where $\Lambda_0$ satisfies
\bqa
\log{\Lambda_0^2\over m_q^2}+F(m_\pi^2)+m_\pi^2F^{\prime}(m_\pi^2)
&=&0\;.
\label{l0}
\eqa

In Appendix B, we derive the relations between the parameters
in the on-shell and $\overline{\rm MS}$ schemes. 
The parameters in the $\overline{\rm MS}$ can then be expressed
in terms of the physical quantities $m_{\sigma}^2$ etc. Evaluating
these para\-meters at $\Lambda=\Lambda_0$ gives $m_0$ etc as functions
of $m_{\sigma}^2$, $m_{\pi}^2$, $m_q$, and $f_{\pi}$. 
For example, evaluating $g_{\ms}^2(\Lambda)$ at
$\Lambda=\Lambda_0$, 
yields $g_0={m_q\over f_{\pi}}$. The other parameters can be
found in the same manner. Inserting the solutions 
(\ref{sol0})-(\ref{sol5}) into (\ref{potms}) and 
expressing the parameters in terms of physical quantities, we finally
obtain the renormalized vacuum energy in the large-$N_c$ limit
\begin{widetext}
\bqa\nonumber
V_{1}&=&
{1\over2}B_0^2\left\{
1+N_c
\sum_f{2q_f^2\over3(4\pi)^2}\log{m_q^2\over2|q_fB|}
\right\}
+{1\over2}f_{\pi}^2q^2
\left\{1-\dfrac{4 m_q^2N_c}{(4\pi)^2f_\pi^2}
\left[\log\mbox{$\Delta^2\over m_q^2$}
+F(m_\pi^2)+m_\pi^2F^{\prime}(m_\pi^2)\right]
\right\}{\Delta^2\over m_q^2}
\\ && \nonumber
+\dfrac{3}{4}m_\pi^2 f_\pi^2
\left\{1-\dfrac{4 m_q^2N_c}{(4\pi)^2f_\pi^2}m_\pi^2F^{\prime}(m_\pi^2)
\right\}\dfrac{\Delta^2}{m_q^2}
\\ \nonumber &&
 -\dfrac{1}{4}m_\sigma^2 f_\pi^2
\left\{
1 +\dfrac{4 m_q^2N_c}{(4\pi)^2f_\pi^2}
\left[ \left(1-\mbox{$4m_q^2\over m_\sigma^2$}
\right)F(m_\sigma^2)
 +\dfrac{4m_q^2}{m_\sigma^2}
-F(m_\pi^2)-m_\pi^2F^{\prime}(m_\pi^2)
\right]\right\}\dfrac{\Delta^2}{m_q^2} \\ \nonumber
 & & + \dfrac{1}{8}m_\sigma^2 f_\pi^2
\left\{ 1 -\dfrac{4 m_q^2  N_c}{(4\pi)^2f_\pi^2}\left[
\dfrac{4m_q^2}{m_\sigma^2}
\left( 
\log\mbox{$\Delta^2\over m_q^2$}
-\mbox{$3\over2$}
\right) -\left( 1 -\mbox{$4m_q^2\over m_\sigma^2$}\right)F(m_\sigma^2)
+F(m_\pi^2)+m_\pi^2F^{\prime}(m_\pi^2)\right]
 \right\}\dfrac{\Delta^4}{m_q^4}
\\ \nonumber&&
- \dfrac{1}{8}m_\pi^2 f_\pi^2
\left[1-\dfrac{4 m_q^2N_c}{(4\pi)^2f_\pi^2}m_\pi^2F^{\prime}(m_\pi^2)\right]
\dfrac{\Delta^4}{m_q^4}
-m_\pi^2f_\pi^2\left[
1-\dfrac{4 m_q^2  N_c}{(4\pi)^2f_\pi^2}m_\pi^2F^{\prime}(m_\pi^2)
\right]\dfrac{\Delta}{m_q}
-{N_cq^4\over6(4\pi)^2}
\\ \nonumber&&
+{N_c\over3(4\pi)^2}
\left[q\sqrt{{q^2\over4}-\Delta^2}(26\Delta^2+q^2)
-12\Delta^2(\Delta^2+q^2)\log{q+2\sqrt{{q^2\over4}-\Delta^2}\over2\Delta}
\right]
\theta(\mbox{$q\over2$}-\Delta)
\\&& 
+
{N_c\over(4\pi)^{3\over2}}\sum_{f,\pm}\int_{p_z}\int_0^{\infty}{ds\over s^{5\over2}}
\bigg\{\Big[
|q_fB|s\coth(|q_fB|s)-1\Big]e^{-s a_{\pm}^2}
-{1\over3}e^{-2s|q_fB|}(q_fBs)^2e^{-s p_z^2}
\bigg\}
\label{potno}
\;.
\eqa
\end{widetext}

\section{Results and discussion}
\label{results}
In the numerical results presented below,
we use a sigma mass of $m_{\sigma}=600$ MeV, a pion mass of
$m_{\pi}=140$ MeV, a quark mass of $m_q=300$ MeV,
and a pion decay constant $f_{\pi}=93$ MeV.

\subsection{Homogeneous case}
\label{homocase}
\begin{figure}[htb]
\includegraphics[width=0.48\textwidth]{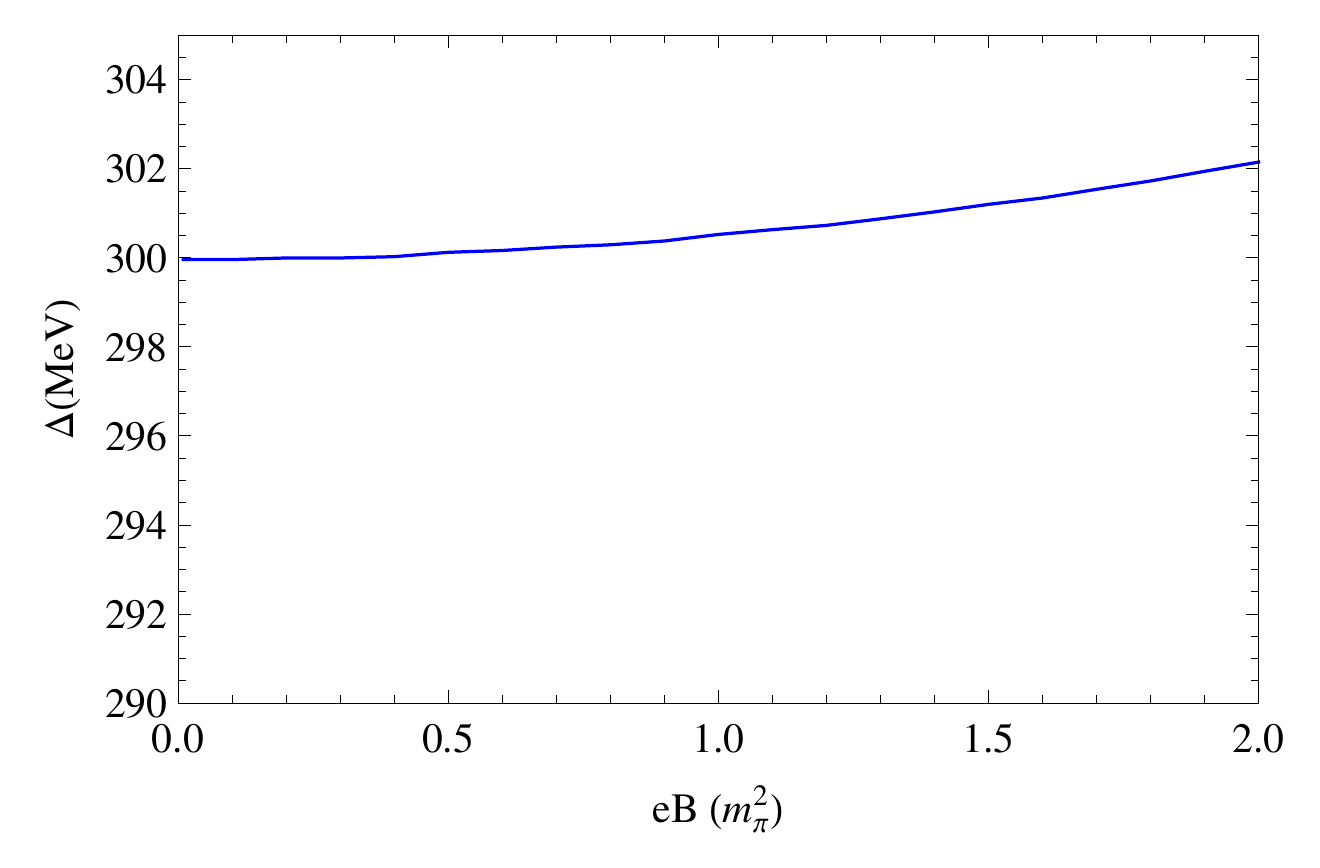}
\caption{$\Delta$ as a function $|eB|$ in units of 
$m_{\pi}^2$ for $\mu=0$.}
\label{delta}
\end{figure}

First we restrict ourselves to a homogeneous condensate.
The vacuum part of the thermodynamic potential is then found by
setting $q=0$ in Eq. (\ref{potno}).
In that case, one can integrate over
$p_z$ and $s$ explicitly, which yields
\begin{widetext}
\bqa\nonumber
V_{1}&=&
{1\over2}B_0^2\left\{
1+N_c
\sum_f{2q_f^2\over3(4\pi)^2}\log{m_q^2\over2|q_fB|}
\right\}
+\dfrac{3}{4}m_\pi^2 f_\pi^2
\left\{1-\dfrac{4 m_q^2N_c}{(4\pi)^2f_\pi^2}m_\pi^2F^{\prime}(m_\pi^2)
\right\}\dfrac{\Delta^2}{m_q^2}
\\ \nonumber &&
 -\dfrac{1}{4}m_\sigma^2 f_\pi^2
\left\{
1 +\dfrac{4 m_q^2N_c}{(4\pi)^2f_\pi^2}
\left[ \left(1-\mbox{$4m_q^2\over m_\sigma^2$}
\right)F(m_\sigma^2)
 +\dfrac{4m_q^2}{m_\sigma^2}
-F(m_\pi^2)-m_\pi^2F^{\prime}(m_\pi^2)
\right]\right\}\dfrac{\Delta^2}{m_q^2} \\ \nonumber
 & & + \dfrac{1}{8}m_\sigma^2 f_\pi^2
\left\{ 1 -\dfrac{4 m_q^2  N_c}{(4\pi)^2f_\pi^2}\left[
\dfrac{4m_q^2}{m_\sigma^2}
\left( 
\log\mbox{$\Delta^2\over m_q^2$}
-\mbox{$3\over2$}
\right) -\left( 1 -\mbox{$4m_q^2\over m_\sigma^2$}\right)F(m_\sigma^2)
+F(m_\pi^2)+m_\pi^2F^{\prime}(m_\pi^2)\right]
 \right\}\dfrac{\Delta^4}{m_q^4}
\\ \nonumber&&
- \dfrac{1}{8}m_\pi^2 f_\pi^2
\left[1-\dfrac{4 m_q^2N_c}{(4\pi)^2f_\pi^2}m_\pi^2F^{\prime}(m_\pi^2)\right]
\dfrac{\Delta^4}{m_q^4}
-m_\pi^2f_\pi^2\left[
1-\dfrac{4 m_q^2  N_c}{(4\pi)^2f_\pi^2}m_\pi^2F^{\prime}(m_\pi^2)
\right]\dfrac{\Delta}{m_q}
-{N_cq^4\over6(4\pi)^2}
\\&& 
-\sum_{f}{8N_c(q_fB)^2\over(4\pi)^2}
\left[
\zeta^{(1,0)}(-1,x_f)+{1\over4}x_f^2-{1\over2}x_f^2\log x_f
+{1\over2}x_f\log x_f-{1\over12}\right]
\label{potno2}
\;,
\eqa
where $x_f={\Delta^2\over2|q_fB|}$, $\zeta(a,x_f)$ is the 
Hurwitz zeta-function, and 
$\zeta^{(1,0)}(-1,x_f)={\partial\zeta(a,x_f)\over\partial a}|_{a=-1}$.

The finite-density contribution $V_{\rm den}$ is given by the 
zero-temperature limit of the logarithmic terms in Eq. (\ref{vacsum}).
In the homogeneous case, it reduces to
\bqa
V_{\rm den}&=&-N_c\sum_{f,k,\sigma_z}{|q_fB|\over2\pi}\int_{p_z}
(\mu-E)\theta(\mu-E)\;,
\eqa
where $E=\sqrt{p_z^2+\Delta^2+M_B^2}$. 
Integrating over $p_z$ from $p_z=0$
to $p_z=p_F=\sqrt{\mu^2-\Delta^2-M_B^2}$, we obtain
\bqa
V_{\rm den}&=&-N_c\sum_{f,k,\sigma_z}{|q_fB|\over4\pi^2}
\Bigg[
\mu\sqrt{\mu^2-\Delta^2-M_B^2}
-\sqrt{\Delta^2+M_B^2}
\log{\mu+\sqrt{\mu^2-\Delta^2-M_B^2}\over\sqrt{\Delta^2+M_B^2}}
\Bigg]
\times\theta(\mu-\sqrt{\Delta^2-M_B^2})
\;.
\label{vden}
\eqa
The quark density is given by
\bqa
\rho&=&-{\partial V_{\rm den}\over\partial\mu}
=N_c\sum_{f,k,\sigma_z}{|q_fB|\over2\pi^2}
\sqrt{\mu^2-\Delta^2-M_B^2}
\theta(\mu-\sqrt{\Delta^2-M_B^2})\;.
\label{denso}
\eqa
\end{widetext}
The sum over Landau levels
is cut off due to the theta function and the highest Landau level 
included in the sum is for each quark flavor given by
\bqa
k_{{\rm max}, f} &=& \dfrac{\mu^2-\Delta^2}{2|q_fB|}\;.
\label{kmaxhom}
\eqa 
In Fig. \ref{delta}, we show $\Delta$ 
as a function of the magnetic field in in units of $m_{\pi}^2$
in the vacuum, i.e, for $\mu=0$.
The quark mass is increasing as a function of $B$, which implies that 
the system shows magnetic catalysis. Magnetic catalysis in the vacuum
is a robust result, which has been found in lattice 
simulations \cite{bali2} as well as model calculations,
see \cite{revmag} for a review.

\begin{figure}[htb]
\includegraphics[width=0.48\textwidth]{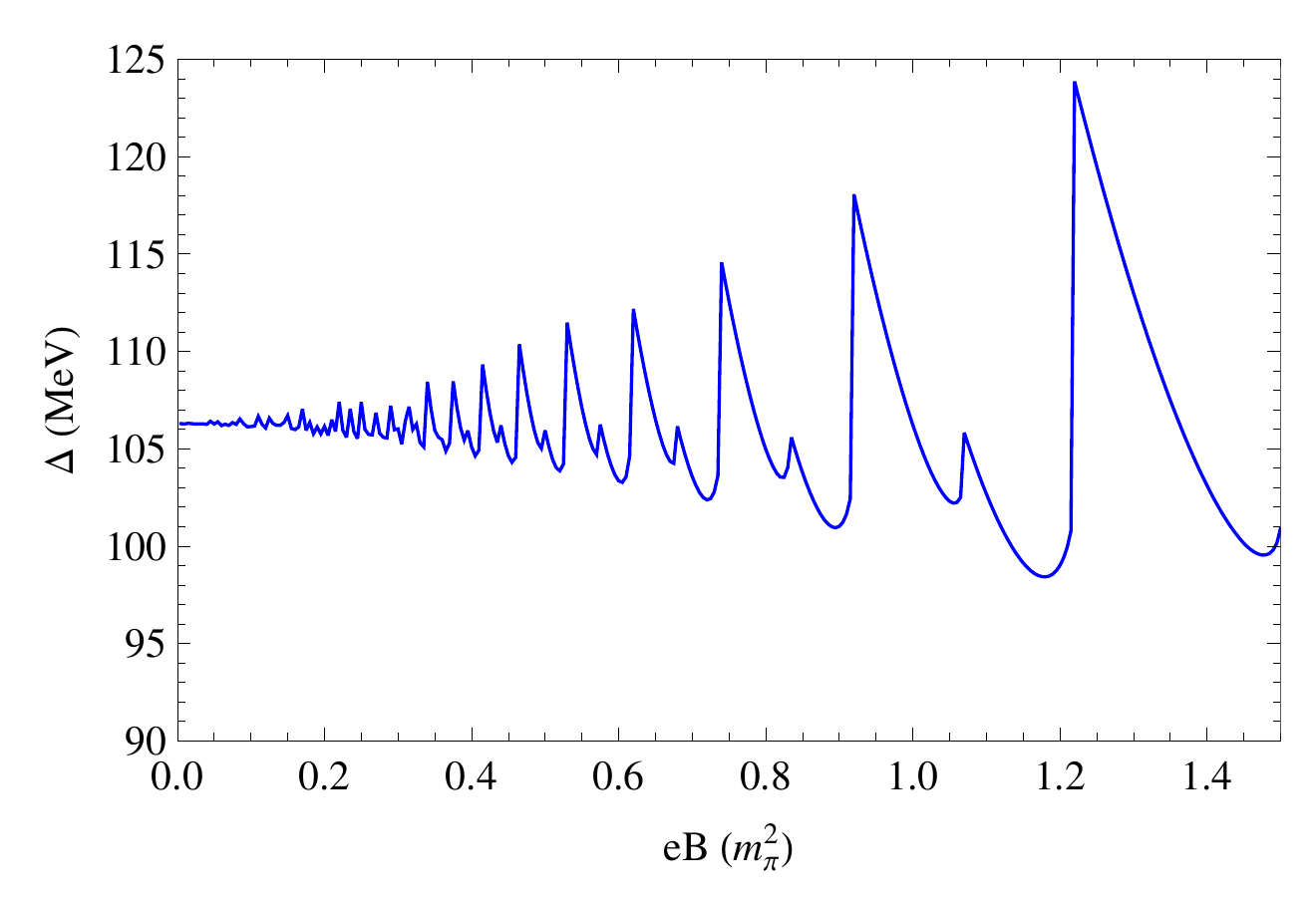}
\caption{$\Delta$ as a function $|eB|$ in units of 
$m_{\pi}^2$ for $\mu=330$ MeV.}
\label{delta0}
\end{figure}

In Fig. \ref{delta0}, we show $\Delta$ as a function of $|eB|$
for $\mu=330$ MeV.
We notice the oscillations in the
parameters $\Delta$ as $|eB|$ increases. Such oscillatory behavior
is known from cold dense systems in an external magnetic field and
caused by the discrete nature of the Landau levels.
For constant chemical potential the maximum number of Landau 
levels that are summed over in (\ref{vden}) decreases with $\frac{1}{|eB|}$. 
The successive large and small jumps in $\Delta$ are related to summing over one 
less Landau level of the up- and down-quark
due to their different electric charge. 
For a large enough magnetic field 
only the lowest Landau level contributes and the oscillations stop.

\begin{figure}[htb]
\includegraphics[width=0.48\textwidth]{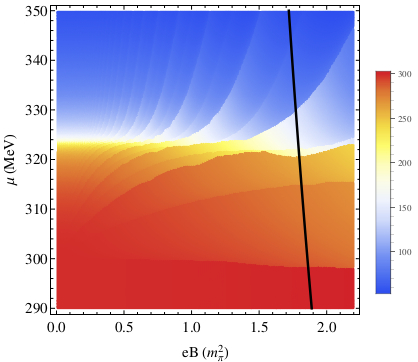}
\caption{The $\mu-B$ phase diagram of the homogeneous model. The black line 
indicated the critical magnetic field as a function of $\mu$. See main text
for details.}
\label{phase}
\end{figure}

In Fig. \ref{phase}, we show the full phase diagram in the 
homogeneous case. One can clearly see the critical lines associated with the 
different Landau levels. The critical chemical potential $\mu_{c,0}$, which 
indicates the transition from the vacuum phase to the one with non-zero quark 
density, becomes lower with increasing magnetic field strength. It has been 
shown in \cite{brauner} that the true ground state of QCD in a magnetic field 
is a chiral soliton lattice, above a critical value of $B$ that depends on 
$\mu_B$. 
This crtical value $B_c(\mu)$ has been 
estimated in Ref. \cite{brauner}
and is shown in
Fig. \ref{phase} as a black line.

At this point, it is appropriate to compare the quark-meson model to
the NJL model and point out some differences.
At zero magnetic field $B$, the NJL model exhibits spontaneous symmetry
breaking only if the coupling $G$ is larger than some critical coupling
$G_c=4\pi^2/\Lambda^2$, where $\Lambda$ is a sharp 
momentum cutoff which is used to regulate the fermionic vacuum fluctuations. 
Thus, in the NJL model vacuum fluctuations induce spontaneous symmetry
breaking, which is in contrast to the 
QM model, as symmetry breaking has been
implemented at tree level using a negative mass parameter $m^2$ in the potential.

Moreover, for $G<G_c$ an arbitrarily small magnetic field 
induces symmetry breaking. This is often referred to as 
dynamical chiral symmetry breaking (DCSB) and was first observed
in the NJL model in 2+1 dimensions \cite{klim21}.
The mechanism behind was later explained in \cite{gusy};
the magnetic field reduces the dyna\-mics of the two spatial
transverse directions, leaving an effectively 1+1 dimensional 
system. This is reminiscent of the formation of a gap in superconductors.
We expect DCSM to be present in the QM model as well but this has
not been studied since in phenomenological applications,
symmetry breaking is implemented at tree level.

\subsection{Inhomogeneous case}
\label{inhomocase}
In the inhomogeneous case, the 
finite-density contribution to the potential reads
\bqa
V_{\rm den}&=&-N_c\sum_{f,\pm,k,\sigma_z}{|q_fB|\over2\pi}\int_{p_z}
(\mu-E_{\pm})\theta(\mu-E_{\pm})\;,
\label{dens2}
\eqa
where $E_{\pm}$ is given by Eq. (\ref{disprel}).
In section \ref{qm}, we showed that the 
vacuum energy is independent of $q$ 
in the limit $\Delta\rightarrow0$, which indicates that
it is meaningful.
Setting $\Delta=0$ in the dispersion relation
(\ref{disprel}) and evaluating the integral over $p_z$
in Eq. (\ref{dens2}), one finds that
$V_{\rm den}$ is independent of $q$ as well.
Thus the full thermodynamic potential given by the sum of Eqs. (\ref{potno})
and (\ref{dens2}) is well defined.

The quark density can be calculated explicitly and reads
\bqa\nonumber
\rho&=&-{\partial V_{\rm den}\over\partial\mu}
\\ \nonumber
&=&N_c\sum_{f,\pm,k,\sigma_z}{|q_fB|\over4\pi^2}\int_{p_z}
\theta(\mu-E_{\pm})
\\ \nonumber
&=&N_c\sum_{f,\pm,k,\sigma_z}^{k_{max}}{|q_fB|\over4\pi^2}
\sqrt{\left(\sqrt{\mu^2-M_B^2}\mp{q\over2}\right)^2-\Delta^2}
\;.
\\ 
\eqa
It reduces to the homogeneous case (\ref{denso}) for $q=0$ as it should. 
The highest Landau level in the sum is 
for each quark flavor given by
\bqa
k^\pm_{{\rm max}, f} &=& 
\dfrac{\mu^2-\left(\Delta \pm \frac{q}{2} \right)^2}{2|q_fB|}\;.
\label{kmaxinhom}
\eqa

\begin{figure}[htb]
\includegraphics[width=0.48\textwidth]{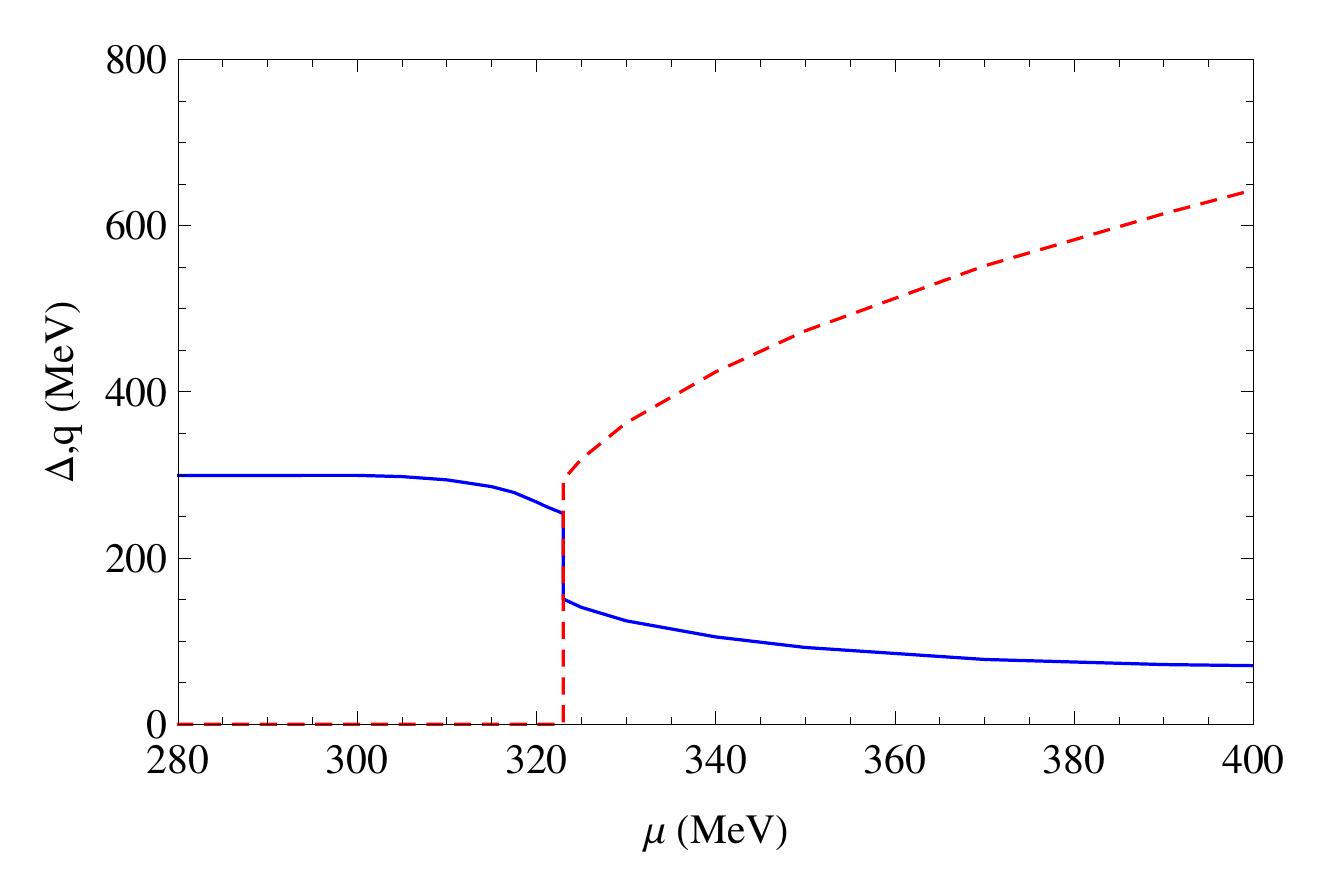}
\caption{ $\Delta$ (blue solid line) and $q$ (red dashed line)
as functions of the chemical potential  $\mu$
for $B=0$.}
\label{b0}
\end{figure}

In Fig. \ref{b0}, we show the magnitude of the chiral condensate $\Delta$ 
(blue solid line)
and the wave vector $q$ (red dashed line)
as functions of $\mu$ for zero magnetic 
field.\footnote{The explicit expressions for the free energy and the quark 
density
for $B=0$ can be found in Ref. \cite{onshell}.}
The vacuum phase extends from $\mu=0$ to $\mu=\mu_{c,0}=300$ MeV. In this
phase, all physical quantities are independent of $\mu$, in particular
$\rho=0$. This is followed by the transition to a homogeneous phase
with a finite quark density. This phase extends to $\mu=\mu_{c,1}=323$ MeV
at which there is a first-order transition to an inhomogeneous phase
with finite quark density and nonzero wave vector $q$.

\begin{figure}[htb]
\includegraphics[width=0.48\textwidth]{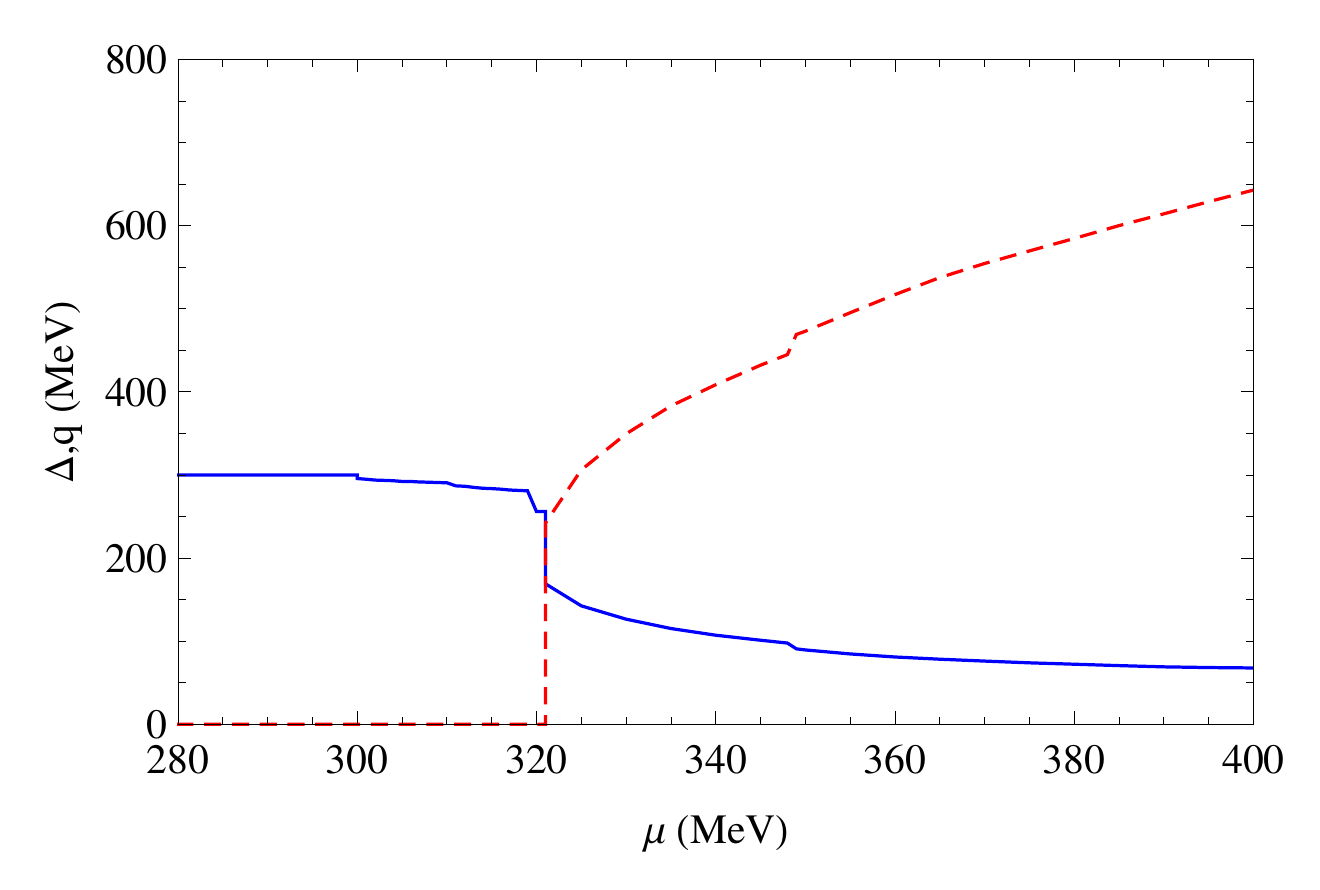}
\caption{$\Delta$ (blue solid line)
and the wave vector $q$ (red dashed line) as functions of $\mu$
for $|eB|=m_{\pi}^2$.}
\label{eb1}
\end{figure}

In Fig. \ref{eb1}, we show the magnitude $\Delta$ (blue solid line)
and the wave vector $q$ (red dashed line) as functions of $\mu$
for $|eB|=m_{\pi}^2$. Again, the vacuum phase exists for 
$\mu=0$ to $\mu=\mu_{c,0}=300$ MeV, whereafter several homogeneous
phases with finite quark density appear.
The inhomogeneous phase starts at $\mu=\mu_{c,1}=321$ MeV. One can clearly
see the successive jumps in the order parameters $\Delta$ and $q$.
For a constant magnetic field the number of relevant Landau 
levels increases with $\mu^2$. When the next Landau level 
is included in the sum, the values of the chiral condensate and wave vector 
jump. It turns out that $\Delta$ starts increasing again past $\mu=415$, and
beyond $\mu=505$ we can no longer find a minimum of the effective
potential. This is not worrisome as one cannot trust the model
for chemical potentials this large anyway.
Finally, 
the effects of a magnetic field on the inhomogeneous phase has 
been studied before in \cite{frolov} using the NJL model. There it has been 
found that for non-zero magnetic field the wave vector increases linearly up to 
the onset of a strong inhomogeneous phase. In contrast to these results we find 
zero wave vector for all $\mu$ smaller than $\mu_{c,1}$.

\begin{figure}[htb]
\includegraphics[width=0.48\textwidth]{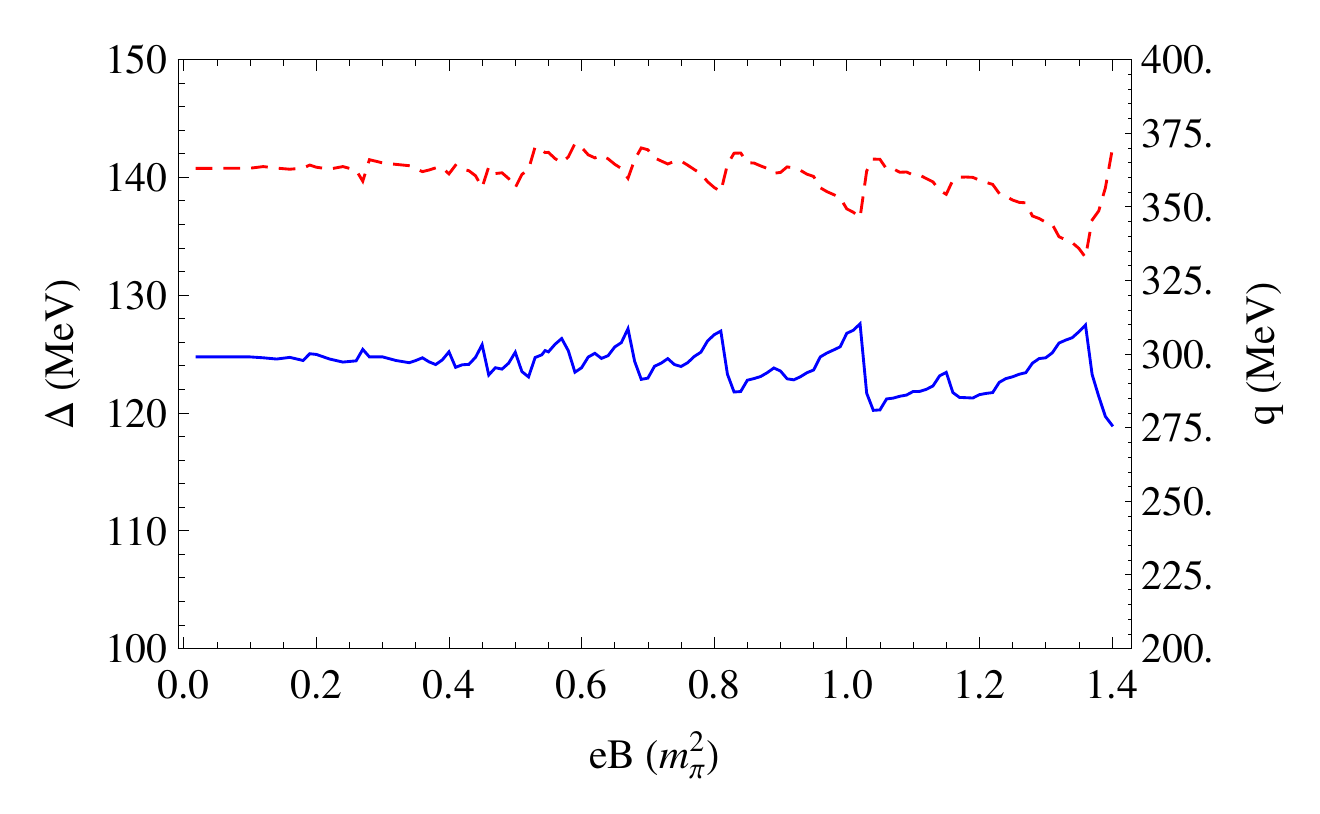}
\caption{$\Delta$ (blue solid line)
and the wave vector $q$ (red dashed line) as functions $|eB|$
for $\mu=330$ MeV.}
\label{330}
\end{figure}

In Fig. \ref{330}, we show 
$\Delta$ (blue solid line)
and the wave vector $q$ (red dashed line) as functions $|eB|$
for $\mu=330$ MeV. We have chosen a value for $\mu$ that lies
in the inhomogeneous phase for $B=0$. 
$\Delta$ is oscillating just like in the
homogeneous case and the value of $\Delta$ for a given $|eB|$
is larger than in the homogeneous case (see Fig. \ref{delta0}). 

The ansatz (\ref{ansatz}) assumes that the wave vector $q$ is parallel to the 
magnetic
field $B$, which is only a special case. The most general case has
the wave vector pointing in an arbitrary direction, allowing for a nonzero
component $q_{\perp}$. In this case the rotational symmetry is completely
broken. Then, however, the spectrum is not known, which prevents us 
from carrying out a complete analysis of the problem. In Ref. \cite{frolov},
the authors used perturbation theory for two nearly degenerate levels
to calculate the correction to the energy levels where the perturbation
is a function of $q_{\perp}$. Using these results, they numerically calculated 
the
second derivative of the thermodynamic potential 
${\partial^2\Omega\over\partial q_{\perp}^2}$ in the
NJL model. 
It turns out that this quantity is positive everywhere, indicating
stability of the thermodynamic potential with respect to $q_{\perp}$.

\section*{Acknowledgments}
The authors would like to thank P. Adhikari and T. Brauner 
for useful discussions.

\appendix
\section{Integrals in dimensional regularization}
In order to calculate the vacuum energy for $B=0$, we need 
a number of integrals in three dimensions. These integrals are
divergent in the ultraviolet and regularized using dimensional regularization.
In analogy with Eq. (\ref{definte}), we define the dimensionally
regulated integral in $d=3-2\epsilon$ dimension as
\bqa\nonumber
\int_{p}&=&
\left({e^{\gamma_E}\Lambda\over4\pi}\right)^{\epsilon}\int{d^{d}p\over(2\pi)^{d}}
\\ &=&
\left({e^{\gamma_E}\Lambda\over4\pi}\right)^{\epsilon}
\int{dp_z\over2\pi}\int{d^{d-1}p_{\perp}\over(2\pi)^{d-1}}
\;,
\label{definte2}
\eqa
where $p_{\perp}^2=p_x^2+p_y^2$.
\begin{widetext}
The specific integrals, we need are
\bqa
\int_p\sqrt{u^2+p^2_{\perp}}
&=&
-\left({e^{\gamma_E}\Lambda^2\over\Delta^2}\right)^{\epsilon}
{\Delta^4\over(4\pi)^2}\Gamma(-2+\epsilon)
=-
\left({\Lambda^2\over\Delta^2}\right)^{\epsilon}
{\Delta^4\over2(4\pi)^2}\left[{1\over\epsilon}+{3\over2}+{\cal O}(\epsilon)
\right]
\;,
\\
\int_p{p_{\perp}^2\over(u^2+p^2_{\perp})^{3\over2}}
&=&
\left({e^{\gamma_E}\Lambda^2\over\Delta^2}\right)^{\epsilon}
{4\Delta^2\over(4\pi)^2}\Gamma(\epsilon)
\left({\Lambda^2\over\Delta^2}\right)^{\epsilon}
={4\Delta^2\over(4\pi)^2}\left[{1\over\epsilon}
+{\cal O}(\epsilon)
\right]\;,
\\ 
\int_p{p_{\perp}^2(4u^2-p_{\perp}^2)\over(u^2+p^2_{\perp})^{7\over2}}
&=&
\left({e^{\gamma_E}\Lambda^2\over\Delta^2}\right)^{\epsilon}
{16\over3(4\pi)^2}(-1+\epsilon)\Gamma(1+\epsilon)
=-{16\over3(4\pi)^2}+{\cal O}(\epsilon)\;,
\eqa
where $u=\sqrt{\Delta^2+p_z^2}$.

We also need some integrals in $D=4-2\epsilon$ dimensions.
Specifically, we need the integrals
\bqa
A(m^2)&=&
\int_p{1\over p^2-m^2}
\label{defa}
={im^2\over(4\pi)^2}\left({\Lambda^2\over m^2}\right)^{\epsilon}
\left[{1\over\epsilon}+1+{\cal O}(\epsilon)\right]\;,
\\ \nonumber
B(p^2)&=&
\int_k{1\over(k^2-m_q^2)[(k+p)^2-m_q^2]}
=
{i\over(4\pi)^2}\left({\Lambda^2\over m_q^2}\right)^{\epsilon}\left[
{1\over\epsilon}+F(p^2)+{\cal O}(\epsilon)
\right]\;,
\label{defb}
\\
B^{\prime}(p^2)&=&{i\over(4\pi)^2}F^{\prime}(p^2)\;,
\eqa
where we have defined the functions
\bqa
F(p^2)&=&
-\int_0^1dx\,\log\left[{p^2\over m_q^2}x(x-1)+1\right]
=
2-2r\,{\arctan}\left(\mbox{$1\over r$}\right)
\;,
\label{fdep}
\\
F^{\prime}(p^2)&=&
{4m_q^2r\over p^2(4m_q^2-p^2)}{\arctan}
\left(\mbox{$1\over r$}\right)
-{1\over p^2}
\;.
\label{defbb}
\eqa
\end{widetext}
\section{Parameter fixing}
In this appendix we use the on-shell renormalization scheme to relate the model 
parameters to physical observables on one-loop level. First we introduce the 
bare parameters
\begin{align*}
m_B^2 &= Z_{m^2}m^2\;, &\lambda_B &= Z_\lambda \lambda\;, & g^2_B &= 
Z_{g^2}g^2, 
\\
h_B &= Z_h h\;,& f^2_{\pi,B} &= Z_{f^2_\pi} f^2_\pi\;,& B^2_B &= Z_A B^2 \\
\sigma_B &= Z^{1/2}_\sigma \sigma\;, &
\pi_B &= Z^{1/2}_\pi \pi\;,& \psi_B &= 
Z^{1/2}_\psi \psi\;,
\end{align*}
where $Z_{\sigma} = 1 + \delta Z_\sigma$ and so on. The renormalization constants 
of the model parameters are given in terms of those of the physical parameters 
and from Eq.(2.7) and (2.8) we find
\bqa
\delta m^2 &=& -\frac{1}{2}\left(\delta m_\sigma^2 -3\delta m_\pi^2\right), \quad 
\\ 
\delta \lambda &=& 3\dfrac{\delta m_\sigma^2-\delta m_\pi^2}{f_\pi^2} -\lambda 
\dfrac{\delta f_\pi^2}{f_\pi^2}\;, \\
\delta g^2 &=& \dfrac{\delta m_q^2}{f_\pi^2} - 
g^2\dfrac{\delta f_\pi^2}{f_\pi^2}\;.
\eqa
Considering that $h=m_\pi^2f_\pi -t$, where $t=0$ is the tree-level tadpole,
we find
\bqa
\delta h = \delta m_\pi^2 f_\pi + \frac{1}{2} m_\pi^2 f_\pi 
\dfrac{\delta f_\pi^2}{f_\pi^2} -\delta t.
\eqa
In the large-$N_c$ limit there are no loop corrections to the quark mass or 
quark field renormalization, therefore $\delta Z_\psi=0$ and $\delta m_q=0$, 
which leads to 
\bqa
\dfrac{\delta g^2}{g^2} = -\dfrac{\delta f_\pi^2}{f_\pi^2}\;.
\eqa
Similarly, in the large-$N_c$ limit, 
the corrections to the pion-quark vertex vanish. This implies
\beq
\dfrac{\delta g^2}{g^2} = -\delta Z_\pi,
\eeq
which allows us to write 
\beq
\dfrac{\delta f_\pi^2}{f_\pi^2} = \delta Z_\pi.
\eeq
In the OS scheme, the renormalized mass of each particle is equal to the 
physical one, which is 
given by the pole of the respective propagator. This gives the OS 
renormalization 
conditions for the sigma and pion masses
\beq
\delta m^2_{\sigma,\pi} = -i\Sigma_{\sigma,\pi}(p^2=m^2_{\sigma,\pi}).
\eeq
In addition, the residues of the propagators at the poles are equal to one, 
which gives 
the renormalization conditions for the fields:
\bqa
\delta Z_{\sigma,\pi} = 
i\dfrac{\partial}{\partial p^2}\Sigma_{\sigma,\pi}(p^2)|_{p^2=m^2_{\sigma,\pi}}
\eqa
The self-energies in the large-$N_c$ limit are
\bqa\nonumber
\Sigma_{\sigma}(p^2) &=& -8g^2N_c \left[A(m_q^2)-\frac{1}{2}(p^2-4m_q^2) B(p^2) 
\right]
\\&&
 + \dfrac{4\lambda g f_\pi N_cm_q}{m_\sigma^2}A(m_q^2)\;,
\\  \nonumber
\Sigma_{\pi}(p^2) &=& -8g^2N_c \left[A(m_q^2)-\frac{1}{2}p^2 B(p^2) \right]
\\ &&
+ \dfrac{4\lambda g f_\pi N_cm_q}{3m_\sigma^2}A(m_q^2)\;,
\eqa
where the last term in both equations is the tadpole contribution. 
Their derivatives, i.e the wave function renormalization counterterms are
\bqa
\delta Z_{\sigma}&=&4ig^2N_c
\left[B(m_\sigma^2)+(m_\sigma^2-4m_q^2)B^{\prime}(m_\sigma^2)\right]\;,\\
\delta Z_{\pi}&=&4ig^2N_c
\left[B(m_\pi^2)+m_\pi^2B^{\prime}(m_\pi^2)\right]\;.
\eqa
The tadpole counterterm is 
determined from the vanishing one-point function and reads
\bqa
\delta t = -8iN_cg^2f_\pi A(m_q^2)\;.
\eqa
\begin{widetext}
With this we find the OS renormalization constants of the model parameters
\bqa \nonumber
\delta m_{\os}^2 &=& 8ig^2 N_c \left[ A(m_q^2) +\frac{1}{4}(m_\sigma^2-4m_q^2) 
B(m_\sigma^2) -\frac{3}{4} m_\pi^2 B(m_\pi^2) \right] \\
 &=& \delta m^2_{\text{div}} +\dfrac{4g^2 N_c}{(4\pi)^2} 
\left\{ m^2\log\frac{\Lambda^2}{m_q^2} -2m_q^2-
\frac{1}{2}\left(m_\sigma^2-4m_q^2\right) F(m_\sigma^2) +\frac{3}{2} 
m_\pi^2F(m_\pi^2)
\right\}\; , \\ \nonumber
\delta\lambda_{\text{\os}} &=& -\dfrac{12ig^2N_c}{f_\pi^2}(m_\sigma^2-4m_q^2)
B(m_\sigma^2) +\dfrac{12ig^2N_c}{f_\pi^2}m_\pi^2B(m_\pi^2) 
-4i\lambda g^2N_c\left[B(m_\pi^2) +m_\pi^2B^{\prime}(m_\pi^2)\right] \\ 
\nonumber
 &=& \delta\lambda_{\text{div}} + \dfrac{12g^2 N_cm_\sigma^2}{(4\pi)^2f_\pi^2}
\left[ \left(1-\dfrac{4m_q^2}{m_\sigma^2}\right)
\left[\log\dfrac{\Lambda^2}{m_q^2} +F(m_\sigma^2) \right]
+\log\frac{\Lambda^2}{m_q^2} +F(m_\pi^2)+m_\pi^2F^{\prime}(m_\pi^2) \right] \\
 && \quad -\dfrac{12g^2 N_cm_\pi^2}{(4\pi)^2f_\pi^2} \left[ 2\log\frac{\Lambda^2}
{m_q^2} +2F(m_\pi^2) +m_\pi^2F^{\prime}(m_\pi^2)\right] \;, \\ \nonumber
\delta g^2_{\text{\os}} 
&=& -4ig^4 N_c\left[B(m_\pi^2)+m_\pi^2B^{\prime}(m_\pi^2)\right] \\
&=& \delta g_{\text{div}}^2 +\dfrac{4g^4 N_c}{(4\pi)^2}
\left[ \log\frac{\Lambda^2}{m_q^2} +F(m_\pi^2) +m_\pi^2F^\prime(m_\pi^2) \right] \;,
 \\ \nonumber
\delta h_{\text{\os}} &=& -2ig^2N_cm_\pi^2f_\pi\left[ B(m_\pi^2)-m_\pi^2B^{\prime}
(m_\pi^2)\right] \\ 
 &=& \delta h_{\text{div}} +\dfrac{2g^2N_cm_\pi^2f_\pi}{(4\pi)^2} \left[
\log\frac{\Lambda^2}{m_q^2} +F(m_\pi^2)-m_\pi^2F^{\prime}(m_\pi^2) \right] \;, \\
\delta Z_{\sigma}^{\text{\os}} 
&=& \delta Z_{\sigma,\text{div}} -\dfrac{4g^2 N_c}{(4\pi)^2}
\left[ \log\frac{\Lambda^2}{m_q^2} +F(m_\sigma^2) +(m_\sigma^2-4m_q^2)
F^\prime(m_\sigma^2) \right] 
\;, \\
\delta Z_{\pi}^{\text{\os}} &=& \delta Z_{\pi,\text{div}} -\dfrac{4g^2N_c}{(4\pi)^2} 
\left[ \log\frac{\Lambda^2}{m_q^2} +F(m_\pi^2) +m_\pi^2F^\prime(m_\pi^2) \right]\;,
\eqa
where we have defined the divergent quantities
\begin{align*}
\delta m^2_{\rm div}&=
{4m^2g^2N_c \over(4\pi)^2\epsilon}
\;,&
\delta\lambda_{\rm div}&=
{8N_c\over(4\pi)^2\epsilon}
\left(\lambda g^2-6g^4
\right)
\;,&
\delta g_{\rm div}^2&=
{4g^4N_c\over(4\pi)^2\epsilon}\;, 
\\
\delta Z_{\sigma,\rm div}&=-{4g^2N_c\over(4\pi)^2\epsilon}
\;,&
\delta Z_{\pi,\rm div}&=-{4g^2N_c\over(4\pi)^2\epsilon}
\;,&
\delta h_{\rm }&=
{2g^2hN_c\over(4\pi)^2\epsilon}\;.
\end{align*}
The photon self-energy in the vacuum is
\bqa
\Pi_{\mu\nu}(p^2)&=&i\left(p^2g_{\mu\nu}-p_{\mu}p_{\nu}\right)\Pi(p^2)\;,
\eqa
where $\Pi(p^2)$ in the large-$N_c$ limit is given by the quark-loop
contribution
\bqa
\Pi(p^2) &= -\dfrac{i8N_c}{(4\pi)^2}\sum\limits_f q_f^2 \left\{
\mbox{\Large$1\over6$}\normalsize
\bigg[\dfrac{1}{\epsilon}+\log\left(\dfrac{\Lambda^2}{m_q^2}\right)\bigg]-
\int\limits_0^1 dx\,x(1-x)\log\left[1-x(1-x)\dfrac{p^2}{m_q^2}\right] \right\}\;.
\eqa
The $B$-field renormalization constant
is then determined by the renormalization condition
\bqa
\delta Z_A^{\os} = -i\Pi(p^2=0) 
 = -\dfrac{4N_c}{3(4\pi)^2}\sum\limits_f q_f^2 \left[ \dfrac{1}{\epsilon} + 
\log\left(\dfrac{\Lambda^2}{m_q^2}\right)\right] \;.
\eqa
Since the bare parameters are independent of the renormalization scheme we can 
immediately find the $\overline{\text{MS}}$ parameters from 
$m^2_{\ms} = m^2+\delta m^2_{\os}-\delta m^2_{\ms}$ 
etc. , which gives
\bqa 
B_{\ms}^2&=&B^2+iN_c\sum_f{4(q_fB)^2\over 3(4\pi)^2}B(0)-\delta B_{\ms}^2
=B^2-N_c\sum_f{4(q_fB)^2\over 3(4\pi)^2}\log{\Lambda^2\over m_q^2}\;,
\\ \nonumber
m^2_{\ms} 
&=& m^2 +8ig^2N_c \left[ A(m_q^2) +\mbox{$1\over4$}(m_\sigma^2-4m_q^2)
B(m_\sigma^2) 
-\mbox{$3\over4$}m_{\pi}^2B(m_{\pi}^2)
\right] -\delta m^2_{\ms} \\
 &=& m^2+
\dfrac{4g^2N_c}{(4\pi)^2} 
\left[m^2\log\mbox{$\Lambda^2\over m_q^2$}
-2m_q^2-{1\over2}\left(m_\sigma^2-4m_q^2\right)F(m_\sigma^2) 
+{3\over2}{m_\pi^2}F(m_\pi^2) 
\right]\;, 
\label{osm1}
\\ \nonumber
\lambda_{\ms} 
&=& \lambda 
-\dfrac{12ig^2N_c}{f_{\pi}^2} (m_\sigma^2-4m_q^2)B(m_\sigma^2) 
+\dfrac{12ig^2N_c}{f_{\pi}^2}m_\pi^2B(m_\pi^2) 
-4i\lambda g^2N_c\left[B(m_{\pi}^2)+m_{\pi}^2B^{\prime}(m_{\pi}^2)\right]
- \delta\lambda_{\ms}
\\ \nonumber &=& 
\lambda
+\bigg\{\dfrac{12g^2N_c}{(4\pi)^2f_\pi^2}
\left[(m_\sigma^2-4m_q^2)\left(
\log\mbox{$\Lambda^2\over m_q^2$} +F(m_\sigma^2)
\right)
+m_\sigma^2\left(
\log\mbox{$\Lambda^2\over m_q^2$}
+F(m_\pi^2)
+m_\pi^2F^{\prime}(m_\pi^2)
\right)
\right.
\\ &&
\left.
-m_\pi^2\left(
2\log\mbox{$\Lambda^2\over m_q^2$}+2F(m_\pi^2)
+F^{\prime}(m_\pi^2)\right)\right]
\bigg\}\;,
\label{osl}
\\ \nonumber
g_{\ms}^2  &= &g^2-4ig^4N_c
\left[B(m_\pi^2) +m_\pi^2B^\prime(m_\pi^2) \right] -\delta g_{\ms}^2 
\\ 
 &=& {m_q^2\over f_{\pi}^2}
\left\{1 + \dfrac{4g^2N_c}{(4\pi)^2}\left[
\log\mbox{$\Lambda^2\over m_q^2$}
+F(m_\pi^2) +m_\pi^2F^\prime(m_\pi^2)
\right]\right\}
\;,
\label{g00}
\\ \nonumber
h_{\ms}&=&h-
2ig^2N_cm_\pi^2f_\pi\left[
B(m_\pi^2)-m_\pi^2B^{\prime}(m_\pi^2)
\right]-\delta h_{\ms}
\\
&=&h+{2g^2hN_cm_\pi^2f_\pi\over(4\pi)^2}
\left[\log\mbox{$\Lambda^2\over m_q^2$}
+F(m_\pi^2)-m_\pi^2F^{\prime}(m_\pi^2)
\right]
\;,
\label{hhh}
\eqa
where the integrals $A(m^2)$ and $B(p^2)$
are defined in Eqs. (\ref{defa})--(\ref{defb}), and 
functions $F(p^2)$ and $F^{\prime}(p^2)$
are defined in Eqs. (\ref{fdep})--(\ref{defbb}).
\end{widetext}


\begin{thebibliography}{*}



\bibitem{revalford}

M. G. Alford, A. Schmitt, K. Rajagopal, T. Sch\"aefer,
Rev. Mod. Phys. {\bf 80}, 1455 (2008).

\bibitem{revhatsuda}
K. Fukushima and T. Hatsuda,
Rept. Prog. Phys. {\bf 74},  014001 (2011).


\bibitem{latt1}
Y. Aoki, Z. Fodor, S. Katz, and K. Szabo, Phys. Lett. B {\bf  643}, 46 (2006).
\bibitem{latt2}
Y. Aoki,  S. Borsanyi,  S. Durr,  Z. Fodor,  S.D. Katz et al., JHEP {\bf 0906}, 
088 (2009),
\bibitem{latt3}
S. Borsanyi et al. (Wuppertal-Budapest Collaboration),
JHEP {\bf 1009}, 073 (2010).
\bibitem{latt4}
A. Bazavov, T. Bhattacharya, M. Cheng, C. DeTar, H. Ding et al., Phys.Rev. D
{\bf 85}, 054503 (2012).

\bibitem{svet}
B.  Svetitsky  and  L.  G.  Yaffe,  
Nucl.  Phys.  B {\bf 210},  423 (1982).

\bibitem{fukk}
K. Fukushima, Phys. Lett. B {\bf 591}, 277 (2004).

 	


\bibitem{ebert2000}
D. Ebert, K. G. Klimenko, M. A. Vdovichenko, and A. S. Vshivtsev,
Phys. Rev. D {\bf 61}, 025005 (2000).



\bibitem{inag}
T. Inagaki, D. Kimura, and T. Murata,
Prog. Theor. Phys. Suppl. {\bf 153}, 321 (2004).

\bibitem{manuel}
E. J. Ferrer, V. de la Incera, C. Manuel,
Phys. Rev. Lett. {\bf 95}, 152002 (2005).

\bibitem{jorge}
J. L. Noronha and I. A. Shovkovy, 
Phys. Rev. D {\bf 76}, 105030 (2007),
Erratum: Phys. Rev. D {\bf 86}, 049901 (2012).

\bibitem{harmen}
K. Fukushima and H. J. Warringa,
Phys. Rev. Lett. {\bf 100}, 032007 (2008).


\bibitem{prov1}
D. P. Menezes, M. B. Pinto, S. S. Avancini,
A. Perez Martinez, and C. Providencia,
Phys. Rev. C {\bf 79}, 035807 (2009).


\bibitem{frolov}
I. E. Frolov, V. Ch. Zhukovsky, and K. G. Klimenko,
Phys. Rev. D {\bf 82}, 076002 (2010).


\bibitem{allen}
P. G. Allen and N. N. Scoccola,
Phys. Rev. D {\bf 88}, 094005 (2013).

	
\bibitem{grun}
A. G. Grunfeld, D. P. Menezes, M. B. Pinto, and N. N. Scoccola,
Phys. Rev. D {\bf 90}, 044024 (2014).




\bibitem{sponmag}
R. Yoshiike, K. Nishiyama, and T. Tatsumi,
Phys. Lett. B {\bf 751}, 123 (2015).
\bibitem{cao}
G. Cao and  A. Huang, 
Phys. Rev. D {\bf 93}, 076007 (2016).
	
\bibitem{nishi}
K. Nishiyama, S. Karasawa, and T. Tatsumi,
Phys. Rev. D {\bf 92}, 036008 (2015).

\bibitem{lif}
T. Tatsumi, K. Nishiyama, and S. Karasawa,
Phys. Lett. B {\bf 743}, 66 (2015).	

\bibitem{delia}
M. D'Elia, S. Mukherjee, and F. Sanfilippo,
Phys. Rev. D {\bf 82}, 051501 (2010).


\bibitem{bali1}
G. S. Bali, F. Bruckmann, G. Endrodi, Z. Fodor, 
S. D. Katz, S. Krieg, A. Schafer, and K. K. Szabo,
JHEP 12 {\bf 02}, 044 (2012).


\bibitem{bali2}
G. S. Bali, F. Bruckmann, G. Endrodi, Z. Fodor, 
S. D. Katz, and A. Schafer,
Phys. Rev. D {\bf 86}, 071502 (2012).

\bibitem{endrodi1}
F. Bruckmann, G. Endrodi, T. G. Kovacs,
JHEP 13 {\bf 04}, 112 (2013).




\bibitem{son}
D. T. Son and  M. A. Stephanov,
Phys. Rev. D {\bf 77}, 014021 (2008).

\bibitem{brauner}
T. Brauner and S. Kadam,
JHEP 17 {\bf 03}, 015 (2017).

\bibitem{klev}S. P. Klevansky, Rev. Mod. Phys. {\bf 64}, 649 (1992).

\bibitem{tomas}
J. O. Andersen and T. Brauner,
Phys. Rev. D {\bf 81}, 096004 (2010).



\bibitem{symren}
D. Ebert, N. V. Gubina, K. G. Klimenko, S. G. Kurbanov, and 
V. Ch. Zhukovsky,
Phys. Rev. D {\bf 84}, 025004 (2011).


\bibitem{jens1}
P. Adhikari and J. O. Andersen, Phys. Rev. D {\bf 95}, 036009 (2017).


\bibitem{buballarev}
M. Buballa and S. Carignano, Prog. Part. Nucl. Phys. {\bf 81}, 39 (2015).


\bibitem{fuji} K. Fujikawa, Phys. Rev. D {\bf 21}, 2848 (1980).

\bibitem{revmag}
I. A. Shovkovy, Lect. Notes Phys. {\bf 871}, 13 (2013).

\bibitem{klim21}
K. G. Klimenko, Theor. Math. {\bf 89}, 1161 (1992);
Z. Phys. C {\bf 54}, 323 (1992);
K. G. Klimenko, A. S. Vshivtsev, and B. V. Magntisky, Nuovo Cimento A
{\bf 107}, 439 (1994).
\bibitem{gusy}V. P. Gusynin, V. A. Miransky, and I. A. Shovkovy,
Phys. Rev. Lett. {\bf 73}, 3499 (1994).


\bibitem{gorbar}
E. V. Gorbar, V. A. Miransky, and I. A. Shovkovy. Phys. Rev. C {\bf 80}
032801 (2009).





  





 	













\bibitem{cohen} T. D. Cohen, Phys. Rev. Lett. {\bf 91}, 222001 (2003).



\bibitem{onshell}
P. Adhikari, J. O. Andersen, and P. Kneschke,
Phys. Rev. D {\bf 95}, 036017 (2017).





\end{thebibliography}
\end{document}